# GHz repetition rate, sub-100-fs Ho:CALGO laser at 2.1 µm with watt-level average power

Weichao Yao,[1,*] Mohsen Khalili,[1] Yicheng Wang,[1] Martin Hoffmann,[1] Marcel van Delden,[2] Thomas Musch,[2] and Clara J. Saraceno[1]

[1]*Photonics and Ultrafast Laser Science, Ruhr Universität Bochum, Universitätsstraße 150, 44801 Bochum, Germany*
[2]*Institute of Electronic Circuits, Ruhr Universität Bochum, Universitätsstraße 150, 44801 Bochum, Germany*
*Corresponding author: weichao.yao@ruhr-uni-bochum.de

We report on a GHz fundamental repetition rate Kerr-lens mode-locked Ho:CALGO laser emitting at 2.1 µm. The laser employs a ring-cavity to increase the fundamental repetition rate to 1.179 GHz and can be made to oscillate in both directions stably with nearly identical performance: for counterclockwise oscillation, it generates 93-fs pulses at 1.68 W of average power, whereas 92 fs and 1.69 W were measured for clockwise operation. Our current results represent the highest average power from a 2-µm GHz oscillator and the first sub-100-fs pulse duration from a Ho-based oscillator.

Mode-locked lasers with GHz repetition rates have a plethora of promising applications for example in clocking, RF referencing and frequency comb spectroscopy [1, 2]. For the latter, a high repetition rate in combination with high average power and low-noise levels enable widely spaced, powerful single comb lines, which is beneficial for increasing signal-to-noise ratio in spectroscopy. GHz femtosecond lasers in many wavelength regions are nowadays successfully applied in time or comb-resolved spectroscopy [3, 4]. In addition, high power GHz repetition rate lasers also showed great potential for material processing to improve the ablation efficiency while maintaining the ablation quality [5].

For all these applications, significant efforts were made to generate GHz ultrashort pulses with high average powers. In this regard, many realizations were done with fiber lasers [6, 7], while bulk lasers can provide an attractive combination of average power and low-noise operation [8]. Most efforts in this direction were realized in the 1-µm wavelength region [3, 4, 9], with a state-of-the-art average power of 6.9 W from a Yb bulk oscillator [9]. Meanwhile, there is an increasing interest in operating at longer wavelengths >1.5 µm for many of these applications. For example, for spectroscopy, to directly address greenhouse gas detection [10], or even >2 µm to efficiently reach the molecular fingerprinting region via nonlinear conversion. To date, most 2-µm GHz lasers are based on Tm fibers and Cr:ZnS bulk lasers. In terms of fiber geometry, scaling the average power from GHz oscillators is still a challenge due to the low gain provided by short fibers, with typical average power is in the mW regime [7]. Therefore, multi-stage amplifiers are needed to achieve watt-level average power, making the laser systems complex, and typically increasing noise level. Mode-locked bulk lasers provide a promising alternative to reach watt-level average power directly from a GHz oscillator. Recently, a SESAM mode-locked Cr:ZnS oscillator was reported with a repetition rate of 2 GHz, an average power of 0.8 W, and a pulse duration of 155 fs [11]. In this laser, a specially designed SESAM is necessary to achieve stable mode-locking. In [12], a Kerr-lens mode-locked Cr:ZnS oscillator with adjustable repetition rate generated 120 mW of average power in 50 fs pulses at 1.2 GHz, and 1.2 W was obtained at 1 GHz with a slightly longer pulse duration of 75 fs, representing thus far the highest average power from a 2-µm GHz oscillator.

$Ho^{3+}$-doped bulk materials are another excellent candidate to achieve ultrashort pulses with watt-level average power in this wavelength region [13, 14]. However, two difficulties have so far prevented significant progress in mode-locked GHz operation of Ho-lasers: on the one hand only very recently host materials that support femtosecond pulse durations with $Ho^{3+}$ were demonstrated [13, 14]; and on the other hand, Ho-based gain materials typically feature millisecond-level fluorescence lifetimes at 2 µm, which constitutes a challenge for stable mode-locking at high repetition rate because of likely Q-switching instabilities [15, 16].

In this work, we demonstrate a Kerr-lens mode-locked Ho:CALGO bulk laser at 1.179 GHz fundamental repetition rate. The maximum average power achieved is 1.69 W with an optical-to-optical efficiency of 12.4%. To the best of our knowledge, this is the first GHz Ho-bulk laser system and represents the highest average power from a GHz oscillator at 2 µm so far. In addition, the pulse duration is as short as 92 fs, benefitting from the large gain bandwidth of the Ho:CALGO crystal, representing the first sub-100-fs pulse duration from a Ho-based oscillator.

Figure 1 shows the experimental setup of our Kerr-lens mode-locked GHz Ho:CALGO laser. We used a 10-mm long *a*-cut 3.1 at.% doped Ho:CALGO crystal as the laser gain medium. Both end surfaces were anti-reflection-coated for the wavelength range of 1900 nm to 2200 nm. To ensure good heat dissipation and avoid condensation on the end faces, the cooling temperature of the crystal was kept at 16 °C. At GHz fundamental repetition rate, the cavity length should be less than 15 cm for a standing wave cavity. However, instead of a standing wave cavity, we used a ring-cavity to increase the repetition rate in resonator that is easier to handle. In this way, the laser only passes through the crystal one time for every round-trip, but the resonator design is relaxed. In addition, the total round-trip group delay dispersion (GDD) can be lower, which is beneficial for achieving shorter pulse durations.

The two concave mirrors M1 and output coupler (OC) have the same radius of curvature (RoC) of −50 mm. M1 has a high-transmission coating at the pump wavelength and exhibits high reflectivity (HR) at the laser wavelength. The OC is HR coated at the pump wavelength and exhibits 0.5% transmission at the laser wavelength. The plane mirror M2 has the same reflectivity at the laser emission range as M1, allowing for single-pass absorption in the laser crystal. To achieve stable soliton mode-locking, the total round-trip GDD was optimized and amounted to -1550 fs$^2$, including -1000 fs$^2$ from the dispersive mirror (DM) and -550 fs$^2$ from the 10-mm Ho:CALGO crystal ($\sigma$-polarization). With this arrangement, the fundamental repetition rate of this ring cavity is around 1.2 GHz. A single mode unpolarized 1940-nm Tm fiber laser was used as the pump source. The calculated pump beam waist has a radius of 55 μm inside the crystal. The maximum pump power was limited to 18 W to avoid thermal damage. The single-pass pump absorption of the crystal was estimated by the difference between the incident pump power and the leaked pump power behind M2.

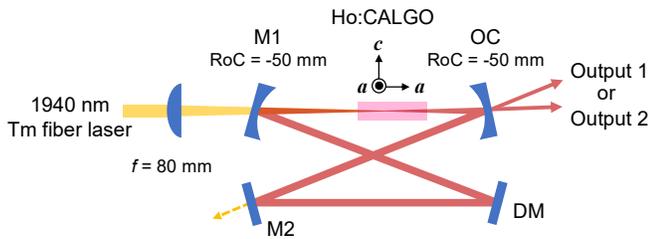

Fig. 1. Experimental setup of the GHz fundamental repetition rate Ho:CALGO laser. M1-2: input mirror; DM: dispersive mirror, -1000 fs$^2$ per bounce; OC: output coupler. The arrows next to the crystal are the crystal axes. Conterclockwise and clockwise outputs are marked as output 1 and output 2, respectively.

The cavity was first aligned in continuous-wave (CW) operation. The OC was aligned to achieve the highest output power for a fixed incident pump power. In this case, the laser works in bidirectional mode, i.e., output 1 and output 2 in Fig. 1 operate simultaneously. Both outputs have nearly the same output power and slope efficiency. A total output power of 3.7 W was obtained at the incident pump power of 18 W (single-pass absorption 61.7%). The natural birefringence of the Ho:CALGO crystal enables a linearly polarized laser output ($\sigma$-polarization) at 2133 nm for both outputs.

To achieve mode-locking, the OC was slightly moved towards the crystal by ~940 μm. In this case, the CW output power decreases rapidly. As a reference, the power of output 1 will reduce from 1.44 W to 0.4 W at 13.6 W of incident pump power. To start mode-locking, the incident pump power was increased to 18 W, then self-starting counterclockwise mode-locking was observed. The unidirectional operation is due to the different Kerr nonlinearities in the crystal for the two opposite directions. In this way, one oscillation direction sees preferential gain and is thus favored [17]. When the laser self-starts (18 W pump power), the laser spectrum exhibits a clear CW breakthrough peak. However, the incident pump power can be safely reduced while maintaining stable and robust mode-locking to operate the laser without CW background.

Figure 2 shows the average output power and pulse duration of the counterclockwise oscillation (output 1) at different incident pump powers. Stable mode-locking can be achieved starting from an incident pump power of 12.4 W (single-pass absorption 60.3%) to a maximum value of 13.6 W (single-pass absorption 59.2%). The corresponding output power ranges from 1.55 W to 1.68 W, resulting in an optical-to-optical efficiency of 12.4% at the maximum output power. In this range, the pulse duration decreases from 99 fs to 93 fs, and the output laser remains linearly polarized in $\sigma$-polarization. When the pump power is further increased, the output power saturates, accompanied by a slight increase in pulse duration. This is because in mode-locking condition, the mode beam radius will shrink below the pump radius at a higher pump power, resulting in slightly mismatch between pump and laser mode, which leads to the saturation of amplitude modulation and the generation of higher-order transverse modes. We confirmed the generation of higher-order transverse modes by the beating frequencies in the radio frequency spectrum [18]. This eventually limits the further increase of intra-cavity intensity and reduction of the pulse duration. At 1.68 W of output power, the output pulses have a fitted full width at half-maximum spectral bandwidth (FWHM, $\Delta\lambda$) of 55 nm at the center wavelength of 2159.1 nm, and a pulse duration (FWHM, $\Delta\tau$) of 93 fs, as shown in Fig. 3(a) and Fig. 3(b), respectively, resulting in a peak power of 13.4 kW. The corresponding time-bandwidth product (TBP) is 0.329, which is close to the TBP (0.315) of an ideal soliton (sech$^2$ fitting). The slight chirp most likely stems from the substrate of the OC and a subsequent collimation system. Further, the ring cavity tends to prevent the generation of satellite pulses [17], we also did not find multi-pulsing in the 16-ps autocorrelation scan, as shown in Fig. 3(b).

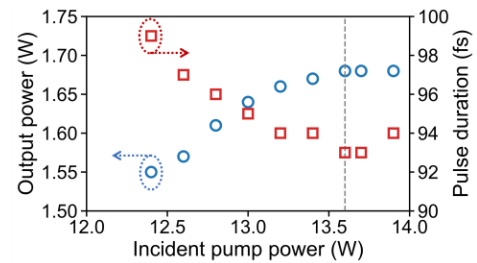

Fig. 2. Average output power and pulse duration of counterclockwise oscillation at different incident pump powers. The dashed line marks the incident pump power for shortest pulse duration in stable mode-locking.

To characterize the mode-locking stability, the radio frequency spectrum was measured with a 12.5-GHz photodiode (EOT-5000, Coherent Inc.) and recorded by a 20-GHz radio frequency analyzer (MS2720T, Anritsu Corp.). The beat notes in Fig. 3(c) exhibit only a slight reduction in intensity due to the limited photodiode bandwidth, and are free of multi-transverse modes beating in a 10-GHz scanning range, indicating stable single-transversal-mode mode-locking [19]. The fundamental repetition rate was measured to be 1179.22 MHz with a spurious free dynamic range of more than 70 dBc, as shown in Fig. 3(d).

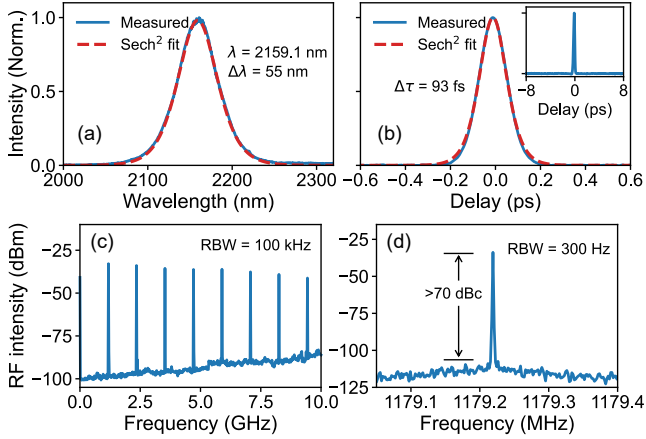

Fig. 3. Laser properties of counterclockwise oscillation at 1.68 W of output power. (a) Laser spectrum; (b) Autocorrelation measurement result, inset is 16-ps autocorrelation scan; (c) Radio frequency spectrum in 10-GHz range. RBW, resolution bandwidth; (d) Radio frequency spectrum of fundamental beat note.

In order to operate the oscillator in clockwise direction (output 2), we adjusted the mirror DM, the positions of the crystal and the OC. The laser exhibits a similar behavior to that of counterclockwise oscillation when it starts the mode-locking, i.e., self-starting in CW at 18 W of incident pump power. Figure 4 shows the average output power and pulse duration at different incident pump power levels in clockwise direction. Only slightly different from the results of counterclockwise oscillation in Fig. 2, the incident pump power for stable mode-locking ranges from 12.8 W (single-pass absorption 58%) to 13.6 W (single-pass absorption 56.9%), corresponding to an average output power of 1.58 W to 1.69 W ($\sigma$-polarization). At the maximum output power, the optical-to-optical efficiency is 12.4%. The change in output power and single-pass absorption is induced by the change of pump-laser mode matching. This occurs because the laser oscillation direction can be adjusted by adjusting the loss and gain of the resonator in one direction [20]. However, multi-transverse modes beating also here is the reason that limits stable mode-locking at a higher pump power. Figure 5 shows the laser properties at 1.69 W of average output power. The central wavelength is 2159.4 nm with a spectral bandwidth of 54.6 nm (FWHM), while the pulse duration is 92 fs (FWHM), resulting in a TBP of 0.323. No multi-pulsing was found in 16-ps autocorrelation scan, as shown in Fig. 5(b). The clean radio frequency spectra in Fig. 5(c) and Fig. 5(d) show stable single transverse-mode mode-locking. The fundamental repetition rate was measured to be 1179.13 MHz with a spurious free dynamic range of more than 70 dBc. A slight reduction in the fundamental repetition rate was induced by the change of cavity length.

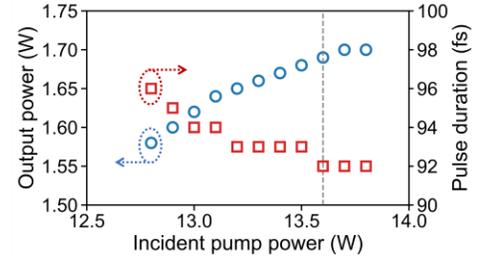

Fig. 4. Average output power and pulse duration of clockwise oscillation at different incident pump power. The dashed line marks the incident pump power for shortest pulse duration in stable mode-locking.

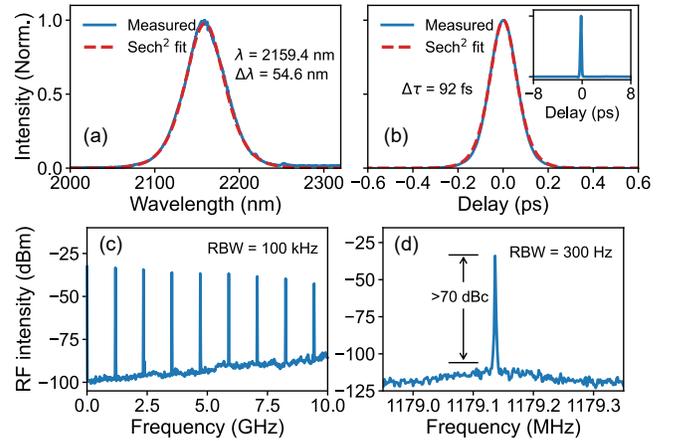

Fig. 5. Laser properties of clockwise oscillation at 1.69 W of average output power. (a) Laser spectrum; (b) Autocorrelation measurement result, inset is 16-ps autocorrelation scan; (c) Radio frequency spectrum in 10-GHz range; (d) Radio frequency spectrum of fundamental beat note.

During our experiment, we did not observe bidirectional mode-locking when we tried to adjust the cavity. In addition, the smaller laser gain of the ring cavity limits us to obtain mode-locking with a higher output transmission, which therefore limits the average output power.

At the maximum output power level, mode-locking in both directions was observed to be stable for hours. Especially, the stability of our GHz laser in clockwise direction was characterized in detail. We measured the average output power at 1.69 W for 1 hour which resulted in an RMS power stability of 0.09%, indicating excellent long-term stability, as shown in Fig. 6(a). Furthermore, the amplitude and phase noises of the laser were measured at a carrier frequency of the first harmonic with a phase noise analyzer (FSWP50, Rohde & Schwarz), as shown in Fig. 6(b) and Fig. 6(c). The noise peaks at low frequency (< 1 kHz) from the amplitude noise relative intensity power spectral density (RIN PSD) should come from mechanical vibration. The weak and broadband noise at an offset frequency of 350 kHz is most likely produced by the relaxation oscillation of the Tm fiber laser we used as a pump [14]. The integrated RIN of the GHz laser is 0.08% in the integration interval from 10 Hz to 10 MHz. Regarding the phase noise, the free-running integrated timing jitter is 13 ps (RMS phase noise, 96 mrad) in the integration interval from 10 Hz to 10 MHz. Affected by a high repetition rate and a low intracavity pulse energy, the phase noise becomes higher at GHz repetition rate, compared with the phase noise of our bulk oscillator at 100-MHz repetition rate [8, 14]. The measured phase noise is comparable to a previously reported GHz-Cr:ZnS laser [11].

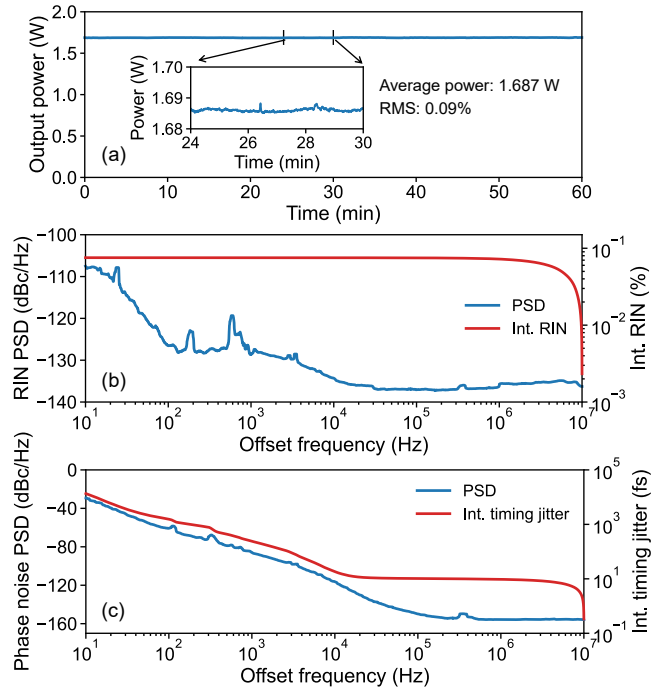

Fig. 6. Stability and noise measurement of GHz Ho:CALGO laser at 1.69 W of average power. (a) Power stability in 1 hour, inset is zoom-in view of output power in 6 minutes; (b) Amplitude noise. RIN: integrated relative intensity. PSD: power spectral density; (c) Phase noise.

In conclusion, we have demonstrated a Kerr-lens mode-locked Ho:CALGO laser with watt-level average power at 1.179 GHz fundamental repetition rate, representing the highest average power GHz oscillator demonstrated to date in the 2-μm wavelength region. The ring laser works in stable and adjustable unidirectional output. For the counterclockwise oscillation, the average power was 1.68 W with 93 fs of pulse duration; while for the clockwise oscillation, the average power was 1.69 W with 92 fs of pulse duration. This also represents the first sub-100-fs Ho laser demonstrated so far. The laser shows excellent power stability and low noise. However, because our mirrors are optimized for the central wavelength of 2100 nm, the difficulty in managing the dispersion at a longer wavelength in our case will limit the generation of a shorter pulse duration. In the near future, we believe significantly shorter pulses in the sub-50 fs regime can be achieved with optimized coatings and Kerr strength. Further power scaling can be considered by better thermal management of the crystal to enable mode-locking with a higher output transmission at higher pump power. The demonstrated laser should be an excellent candidate for repetition frequency and carrier-envelope offset frequency stabilization to achieve a stable GHz frequency comb. Furthermore, the current laser system offers also interesting perspectives for other applications, for example in material processing.

**Funding.** European Research Council (805202, 101138967); Deutsche Forschungsgemeinschaft (390677874, 287022738 TRR 196).

**Acknowledgments.** This project was funded by the Deutsche Forschungsgemeinschaft (DFG) under Germany's Excellence Strategy - EXC 2033 - 390677874 – RESOLV and also under Project-ID 287022738 TRR 196 (SFB/TRR MARIE). These results are part of a project that has received funding from the European Research Council (ERC) under the European Union's Horizon 2020 research and innovation programme (grant agreement No. 805202 - Project Teraqua) and HORIZON-ERC-POC programme (Project 101138967 - Giga2u). W. Yao acknowledges financial support from the Alexander von Humboldt Foundation through a Humboldt Research Fellowship.

**Disclosures.** The authors declare no conflicts of interest.

**Data availability.** Data underlying the results presented in this paper are not publicly available at this time but may be obtained from the authors upon reasonable request.